\documentclass[submission,copyright,creativecommons]{eptcs}
\providecommand{\event}{ESSS 2014} 
\usepackage{breakurl}             
\usepackage{graphicx, color}
\usepackage{latexsym}
\usepackage{amsmath,amssymb}
\usepackage{wrapfig}

\title{Evaluation of A Resilience Embedded System Using Probabilistic Model-Checking}
\author{Ling Fang, Yoriyuki Yamagata, Yutaka Oiwa 
\institute{National Institute of Advanced Industrial Science and Technology, Japan}
\email{$\{$fang-ling, yoriyuki.yamagata, y.oiwa$\}$@aist.go.jp}
}
\def\titlerunning{Evaluation of A Resilience Embedded System Using Probabilistic 
 Model-Checking}
\def\authorrunning{Ling Fang, Yoriyuki Yamagata, Yutaka Oiwa}
\begin{document}
\maketitle

\begin{abstract}
If a Micro Processor Unit (MPU) receives an external electric signal as noise, the system function will freeze or malfunction easily. A new resilience strategy is implemented in order to reset the MPU automatically and stop the MPU from freezing or malfunctioning. The technique is useful for embedded systems which work in non-human environments. However, evaluating resilience strategies is difficult because their effectiveness depends on numerous, complex, interacting factors. 

In this paper, we use probabilistic model checking to evaluate the embedded systems installed with the above mentioned new resilience strategy. Qualitative evaluations are implemented with 6 PCTL formulas, and quantitative evaluations use two kinds of evaluation.  One is system failure reduction, and the other is 
ADT (Average Down Time), the industry standard. Our work demonstrates the benefits brought by the resilience strategy. Experimental results indicate that our evaluation is cost-effective and reliable.

\end{abstract}

\section {Introduction}

\begin{figure}[htbp]
\includegraphics[height=4cm,width=16cm]{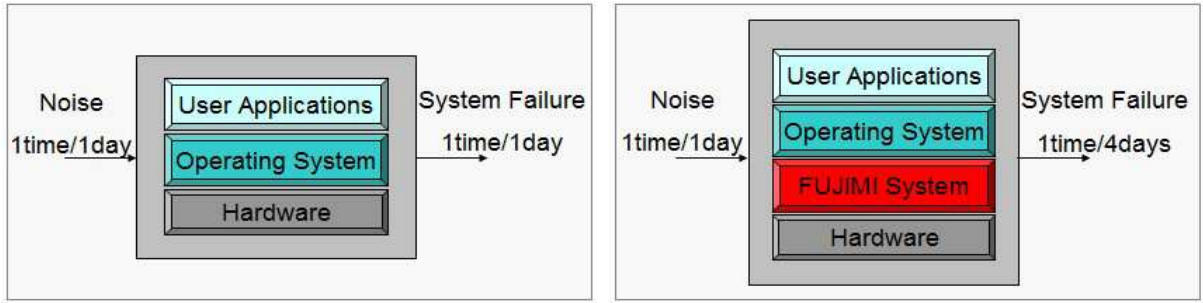}
\caption{The Embedded System With (right) And Without (left) The FUJIMI system}
\label{fig:image1}
\end{figure}

If a Micro Processor Unit (MPU) receives an external electric signal as noise, the system function will freeze or malfunction easily \cite{noise}. Therefore, a resilience strategy for the MPU in a non-human environment is needed so that the embedded system can continue working \cite{resi}. This paper considers a new resilience technique in order to reset the MPU automatically and to prevent the MPU from freezing. 
The application programs restart from backup data. Therefore, the temporal failure has no apparent effect on the application programs. 
This resilience strategy is implemented as a system that works between the hardware and operating system. This system is named FUJIMI\footnote{FUJIMI means always alive  in Japanese.}. 
In the FUJIMI system, Non-Maskable Interrupts (NMI) and Reset signals (RST) are triggered \cite{ARM} periodically to save data for backup and then the backup data is used to reset the MPU. 
Figure \ref{fig:image1} is an example comparing two embedded systems with  and without the FUJIMI system installed to show that system failure is reduced from once a day to once every 4 days.

Evaluating such resilience strategies is a challenging task because the evaluations depend on many complex factors including probabilistic noise environments. 
As a result, empirical experimentation is often a more useful approach for evaluation than proofs or formal approaches. 
However, evaluating a live system under noise attack is time-intensive and costly, and only achieves results under limited cases. This paper presents a formal approach to providing cost-effective and reliable evaluations of resilience strategies by means of simulation and probabilistic model checking. 
We simulate the resilience strategy with a probabilistic model and present a qualitative verification and quantitative assessment for the model. We use PRISM \cite{PRISM} \cite{Hinton}, which is a tool for formal modeling and analysis of systems that exhibit probabilistic behavior. 
DTMC (Discrete-Time Markov Chain) is the specification language, and PCTL  (Probabilistic Computation Tree Logic) is used for analysis \cite{Baier}. 

The results of qualitative verification support certain claims and expectations of resilience strategy. 
Quantitative assessments provide two types of evidence to demonstrate how the resilience strategy improves the system effectiveness: reduction of system failure and standard industry criteria ADT. In other words, the qualitative verification ensures the analysis results from the model are correct and the quantitative assessments demonstrate the benefits of the resilience strategy. 
Moreover, the system can be optimized for different noise environments and strategy variations.  
 The evaluations were difficult to acquire when testing with a fully implemented embedded system until our work.  

The remainder of this paper is structured as follows: 
Section \ref{checker} is a preliminary explanation for probabilistic model checking and PCTL.
Section  \ref{sec:acr} explains the principle and the resilience strategy of the FUJIMI system.  
Section \ref{Modeling} discusses how to construct the formal probabilistic model. 
Section \ref{analysis} describes our methods for system evaluation. 
Section \ref{related} reviews several examples of related work. 
In Section \ref{contribution}, the paper concludes with a discussion about our contributions and future research.

\section{Probabilistic model checking and PCTL}
\label{checker} 

Probabilistic model checking is a formal verification technique for establishing the correctness, performance, and reliability of systems which exhibit stochastic and probabilistic behavior.
As in conventional model checking,
a model of the probabilistic system, usually some variant of a Markov chain, is built and then subjected to algorithmic analysis to establish whether it satisfies a given specification. The specifications are usually stated as formula of probabilistic temporal logic which, in addition to conventional modalities, may include probabilistic operators whose outcome is true/false depending on the probability of certain executions.  

In this paper, we report on our experiences using the probabilistic model checker 
PRISM \cite{PRISM} \cite{Hinton}. 
We use DTMC as the modeling language and use PTCL to verify the properties  \cite{Baier}. 

\subsection{ DTMC (Discrete-Time Markov Chains) Model}
\label{DTMC}

DTMCs are defined as state-transition systems augmented with probabilities. States represent possible configurations of a system. Transitions among states occur at a discrete time and have an associated probability. DTMCs are discrete stochastic processes with the Markov property \cite{PRISM} \cite{Baier}, according to which the probability distribution of future states depend only upon the current state. 
Formally, a labeled DTMC is tuple $M=(S, S_0, P,L)$:  
\begin{itemize}
\item $S$ is a finite set of states
\item $S_0 \in S$ is a set of initial states
\item $P:S\times S \rightarrow [0, 1]$ is a transition probability  matrix where 
$ \sum_{s^{\prime} \in S} P(s, s^{\prime}) = 1$ for all $s\in S$. 
\item $L:S \rightarrow 2^{AP}$ is a labeling function which assigns to each state a set of $Atomic\ Propositions$ which are true in that state.
\end{itemize}

A DTMC model is described as a number of modules which can interact with each other. 
After definition of local and global variables, the behavior of each module is described by a set of commands with the form: $[] guard -> prob_1 : update_1 + ... + prob_n : update_n$.
The guard is a predicate over all the variables in the model. 
Each transition represents an update which the module can make if the guard is true. 
The update is assigned a probability. 

\subsection{ PCTL and Probabilistic Model Checking}
PCTL is a logic language inspired by CTL  \cite{Baier}. 
In place of the existential and universal quantification of CTL, 
PCTL provides the probabilistic operator 
$\wp_{\Join p}(.)$, 
where $p \in [0, 1]$ 
is  probability bound and 
$\Join \in \{ \leq, <, \geq, >  \}$. 
PCTL is defined by the following syntax:

$\Phi ::= true\ \mid a \mid \Phi \wedge \Phi \mid \neg \Phi \mid \wp_{\Join p}(\Psi)$ 

$\Psi ::= X \Phi \mid \Phi U \Phi \mid \Phi U^{\leq t} \Phi$ 

\noindent
Formulas $\Phi$ are named $state\ formulas$ and can be evaluated
over a Boolean domain ($true$, $false$) in each state. Formulas $\Psi$ are named path formulas, and describe a pattern
over the set of all possible paths originating in the state
where they are evaluated. 
The satisfaction relation for a state $s$ in PCTL is:

$s \models true$

$s \models a$ iff $a \in L(s)$

$s \models \neg \Phi$ iff $s\ \models \hspace{-.97em}/\ \Phi$

$s \models \Phi_1 \wedge \Phi_2$ iff $s \models \Phi_1\ and\ s \models \Phi_2$

$s \models \wp_{\Join p}(\Psi)$ iff $Pr(s \models \psi) \Join p$

\noindent
The satisfaction relation for a path formula with respect to a path $\pi$ originating in 
$s\  (i.e.,\pi[0] = s)$ is defined as:

$\pi \models X \Phi$ iff $\pi[1] \models \Phi$

$\pi \models \Phi U \Psi$ iff $\exists j \geq0. (\pi[j] \models \Psi \wedge (\forall k: 0 \leq k <j. \pi[k] \models \Phi))$

$\pi \models \Phi U^{\leq t} {\Psi}$ iff $\exists 0 \leq j \leq t. (\pi[j] \models \Psi \wedge (\forall k: 0 \leq k <j. \pi[k] \models \Phi))$

PCTL is an expressive language that allows many interesting reliability-related properties to be specified. 
 The most important case is a reachability property.
A reachability property states that a state where a certain
characteristic property holds is eventually reached from a
given initial state. Such state may represent a failure state,
in which a transaction executed by the system modeled by
the DTMC eventually terminates, or a success state. 

There are also operators as $S$ and $R$, which are not in the syntax but are well defined in the PRISM for analysis \cite{PRISM}\cite{Ste94}. 
The $S$ operator is used to reason about the steady-state behavior of a model. 
The $R$ operator supports the specification and analysis of properties based on costs and rewards. 
Probabilistic model checking is a decision problem:  
Given a finite Markov chain $M$, and PCTL state formula $\Phi$, 
determine whether $s \models_M \Phi$. 
The following shows several examples for the reliability requirements: 

\begin{itemize}
\item $P_{\ge 0.001}(1 \le s \le 2\ U\ s = 3)$: 
The probability that the system stays between states $1$ and $2$ 
until it reaches state $3$, is greater than 0.001
\item $P_{\ge 0.001}(1 \le s \le 3\ U^{\le5}\ s = 4)$: 
The probability that the system stays between states $1$ and $3$ 
until it reaches state $4$ within 5 steps, is greater than 0.001
\item $S<0.05 [ size /max > 0.75 ]$: 
The long-run probability of the value of $size/max$ being more than $75\%$, is less than 0.05 
\end{itemize}

\section {The FUJIMI System}
\label{sec:acr}  

This section introduces the principle and the resilience strategy of the FUJIMI system. The FUJIMI system is mainly used in game machine currently.  
The Non-stop property of game machine is enforced by the Japanese Government to avoid an argument between the players and the operator if a game machine freezes\footnote{The operation of FUJIMI system has already been proven over 30 million game machines.  
Dragonchip Inc., in Hong Kong, for example, has made one series of FUJIMI MPU. }. 
Before the explanation of the FUJIMI system, we distinguish three kinds of initial process: 

\begin{itemize}

\item $Cold\ start$: The process initializes the hardware, operating system and the user application when the power is on. In general, the process consumes much time.
\item $Hot\ start$: The system initializes the operating system and user application, without initial processing of the hardware. The process needs less time than $cold\ start$, but the intermediate results of the user application are lost.
\item $FUJIMI\ reset$: The periodical reset process for the resilience strategy of FUJIMI system. The process needs less time than $cold\ start$ and $hot\ start$, and the intermediate results of the user application can be preserved.
\end{itemize}

\subsection{System Principle}
For an embedded system which works in an environment without operators present but must work continually, such as a wind velocity meter in open country, keeping the system working continually is an important issue. 
However, it is not rare for microcomputers of embedded system to fall into failures such as $stop$, $freeze$, and $lockup$, which are caused by many problems including illegal instruction, bus errors, etc. 
For example, the instruction pointer may read a data area as an instruction and give the wrong results, or the synchronizations may deviate within the plurality sequencers. 
Noise, such as an abnormal electrical shock in the circuit of the semiconductor is one general reason for these problems. These problems may cause the system to enter into an incorrect status temporarily and this will bring serious effects to the system.
 
Beside well-known methods to recover the system such as watch-dog, power reset and redundant hardware, there are also many resilience strategies such as  DRB (Distributed Recovery Blocks), NVP (N-version programming) and NSCP (N self-checking Programming)  \cite{Joanne}.
However, there is no omnipotent solution for malfunction and freezing of the micro-computer. 
The MPU malfunction can be categorized as below ($level_0$ is a normal status)  \cite{fujimi}: 

\begin{itemize}
 \item  $level_0$: Running normally without any problems
 \item  $level_1$: Spurious Interrupt is caused
 \item  $level_2$: One byte contents of RAM is corrupted or I/O direction is changed
 \item  $level_3$: Many contents of RAM are destroyed
 \item  $level_4$: CPU execution is into the abnormal condition
 \item  $level_5$: A virtual SCR (Silicon Controlled Rectifier) is triggered and active the latch up \cite{renesas}
\end{itemize}

From $level_0$ to $level_3$, the errors can be fixed with the current technologies \cite{ARM}. Once the violation condition on $level_4$ occurs, the MPU cannot resume by itself and must be rescued by a watch-dog in traditional systems. However, the watch-dog can rescue the embedded system only if the problems are detected.
For $level_5$, there is no other way to escape from this except by using Power Off.

A resilience system named FUJIMI \cite{fujimi} is proposed to recover a system from temporary software failure at $level_4$.
The FUJIMI system runs on a general purpose CPU such as M68030, which works between the hardware and the operating system. As shown is Figure \ref{fig:sys}, the clock periodically generates a Non-Maskable Interrupt ($NMI$) and after some interval, generates a reset interrupt ($RST$) to the MPU. The intervals between two $RST$ is called a cycle in this paper.   
When the FUJIMI system receives $NMI$, it saves backup data consisting of 
minimum information for recovery, and then enters into the waiting status. 
When the interrupt $RST$ is signaled,  
the embedded system is reset with the valid backup data, 
or initializes a $hot\ start$ when there is no valid backup data and the system is forced to execute from the beginning of an application.  
The FUJIMI resilience strategy has the following advantages:

\begin{center}
\begin{figure}[t]
\includegraphics[height=4.5cm,width=15cm]{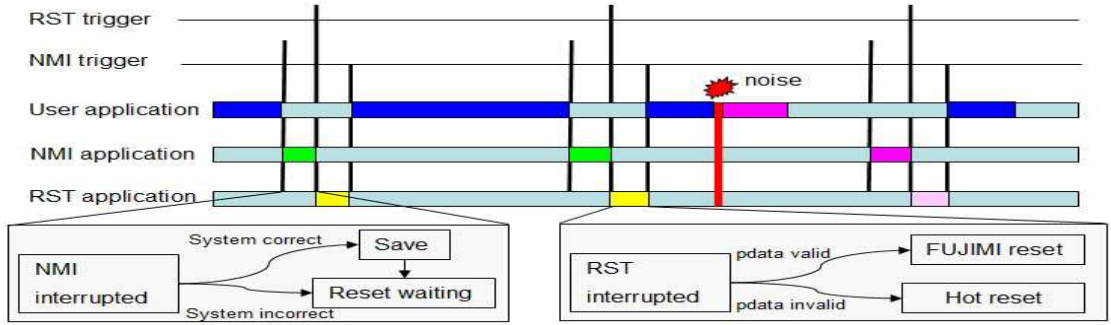}
\caption{The System Principle}
\label{fig:sys}
\end{figure}
\end{center}

\begin{itemize}
\item The strategy recovers systems much faster than general resilience strategies such as watch-dog, which is essential for time-response critical systems. Temporal system failures can be resumed by a short periodical cycle process.
\item The applications can be restarted without loss of the intermediate results. 
The temporal failure caused by noise has no apparent effect on the application programs. 
\item The resilience strategy is implemented without redundancy hardware.  This strategy is useful for a system that has limitation of power or cost.    
\end{itemize}

\subsection{Resilience Strategy}
As previously discussed, the system uses two kinds of signal: $NMI$ and $RST$. 
The basic idea lies in periodically backing-up the data with $NMI$ signal and recover with $RST$ signal. 
We denote the current data as $cdata$, and denote the backup data as $pdata$. 
The data for recovery are saved $n$ times as $pdata_1, pdata_2, \dots, pdata_n$. 
Each recovery data can be reused $m$ times, 
and keeps $valid$ until it is used for $m$ times, where it becomes $invalid$ which represents that it cannot be used anymore.

\begin{figure}[htbp]
\begin{tabular}{cc}
\begin{minipage}{0.35\hsize}
\begin{center}
\includegraphics[height=3.2cm,width=6cm]{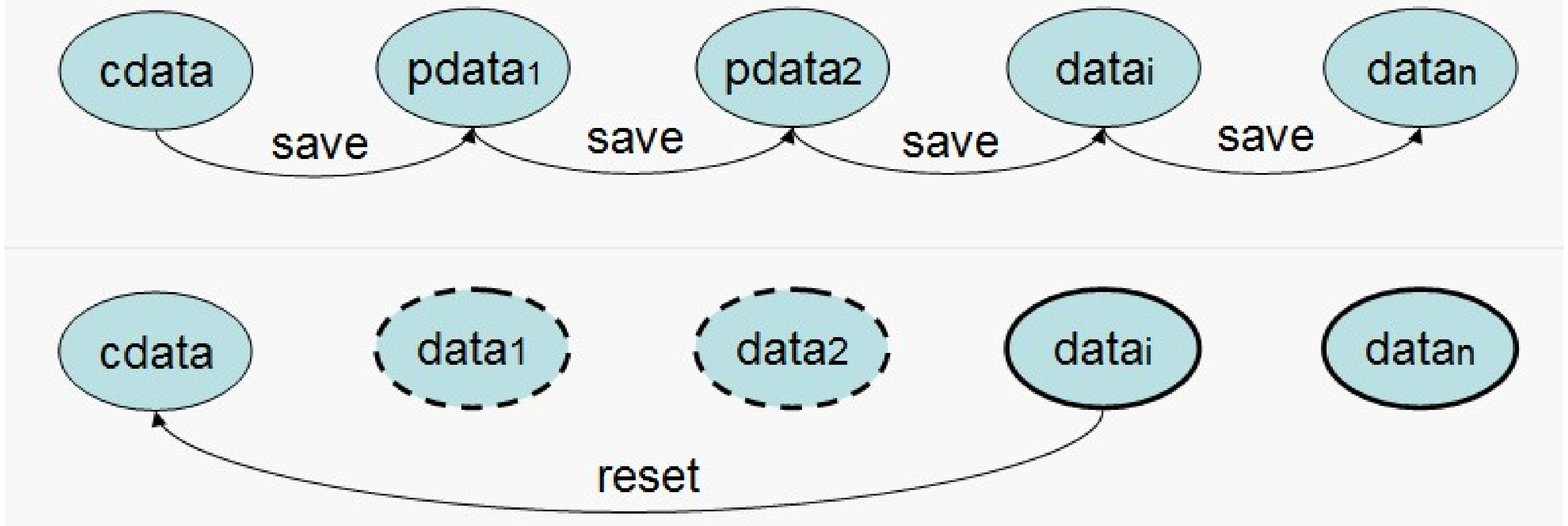}
\caption{The Save and Reset Process}
\label{fig:sr}
\end{center}
\end{minipage}
\begin{minipage}{0.65\hsize}
\begin{center}
\includegraphics[height=5cm,width=9.6cm]{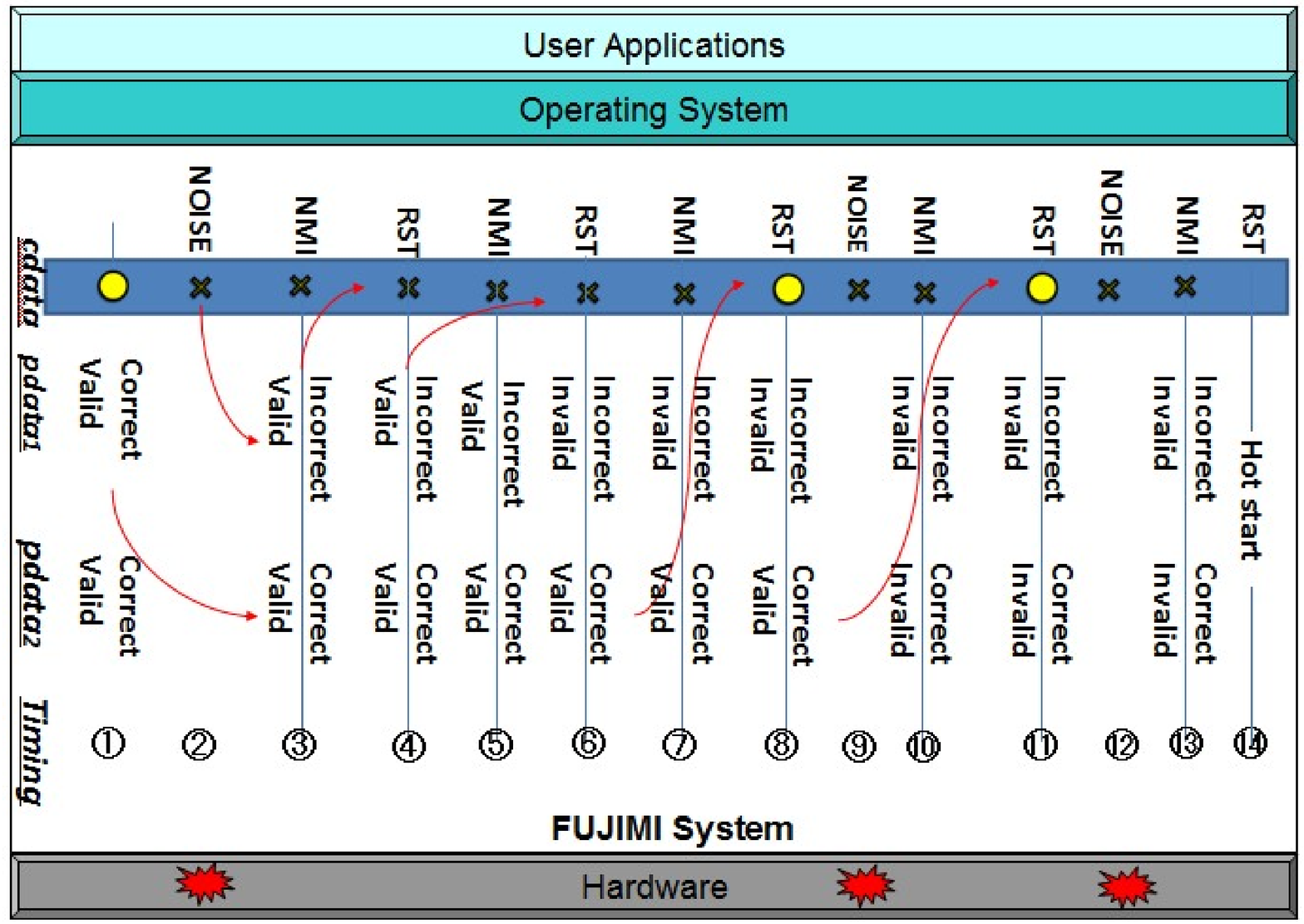}
\caption{The Recover Process}
\label{fig:rcv}
\end{center}
\end{minipage}
\end{tabular}
\end{figure}

When $NMI$ signal arises, 
the FUJIMI system checks whether $CPU$ is in the normal state 
by checking the program counter and the stack pointer. 
As in the upper part of Figure \ref{fig:sr},  if the $MPU$ is in the normal state, 
the save process is carried out from right to left 
as $pdata_{n-1}$ to $pdata_n$,  $pdata_{n-2}$ to $pdata_{n-1}$,  $\dots, pdata_2$ to $pdata_1$, and finally $cdata$ to $pdata_1$. All the $pdata$ are marked as $valid$ 
and the initial number $m$ is assigned to the usage times. 
If the CPU is not in the normal state, the $NMI$ interrupt 
is ignored and the system starts an undisturbed loop until an $RST$ interrupt arises.

When the FUJIMI system receives $RST$ signal, the reset exception handler checks 
the contents of RAM.
If the FUJIMI system is not in the procedure for $cold\ start$ or $hot\ start$, 
the most recent valid $pdata$ is copied to $cdata$ 
as in the lower part of Figure \ref{fig:sr}  (the actual line represents that 
the data is valid and the dotted line represents invalid data).  
For example, if the $pdata_1$ is $valid$, 
then the CPU state is recovered from $pdata_1$,   the usage time will becomes $m-1$. 
If $pdata_1$ has been used for $m$ time and became $invalid$, 
the system will try to recover from the $pdata_2$. 
The process is similar until $pdata_n$ is exhausted and becomes $invalid$.  
When all the $pdata$ are $invalid$, the system cannot be recovered and then the 
$hot\ start$ begins.
In this manner, this system achieves $m*n$ times of software redundancy without overage cost and power consumption caused by hardware redundancy.  

Figure \ref{fig:rcv} shows an example of the recovery process with $2$ previous data 
$pdata_1$ and $pdata_2$, and each $pdata$ can be reused $2$ times. 
The process recovers $2$ noise effectiveness within $4$ cycles. 
However, the system cannot be rescued when the $3^{rd}$ noise occurs, where both  the two $pdata$ are $invalid$, and then a $hot\ start$ begins. 
The following steps from 1-14 explain the process of recovery. 
The number for the items corresponds to the timing shown with the underlining in Figure \ref{fig:rcv}. 

\begin{enumerate}
\item After the $cold\ start$ finishes normally, the initial system status in which the $cdata$, $pdata_1$ and $pdata_2$  are correct, and both $pdata_1$ and $pdata_2$ are $valid$.
\item Noise occurs  for the $1^{st}$ time and makes the $cdata$ incorrect.
\item $NMI$ is triggered.    
The $cdata$ is incorrect and detection fails;   
the correct $pdata_1$ is saved to $pdata_2$, and
the incorrect $cdata$ is saved to $pdata_1$. 
\item The incorrect $pdata_1$ is reset to $cdata$, and the usage times decrease from 2 to 1.
\item The incorrect $cdata$ is detected, and the save process of $NMI$ fails.
\item The incorrect $pdata_1$ is reset to $cdata$, and then $pdata_1$ becomes $invalid$ after decrementing by one for the second time.
\item The incorrect $cdata$ is detected, and the save process of $NMI$ fails.
\item The correct $pdata_2$ is reset to $cdata$, and then the usage times decreases from 2 to 1 because $pdata_1$ is invalid.  The system is recovered for the $1^{st}$ noise.
\item Noise occurs again and makes the $cdata$ incorrect.
\item The incorrect $cdata$ is detected, and the save process of $NMI$ fails.
\item The correct $pdata_2$ is reset to $cdata$, and then becomes $invalid$ after minus $1$ for the $2^{nd}$ time. 
The system is recovered for the $2^{nd}$ noise. 
\item Noise occurs $3^{rd}$ times and makes the $cdata$ incorrect; 
no more valid $pdata$ can be used for recovery because $pdata_1$ and $pdata_2$ are both $invalid$.
\item The incorrect $cdata$ is detected, and the save process of $NMI$ fails.
\item The $hot\ start$ process begins.
\end{enumerate} 

As previously noted, evaluating resilience against noise is another important theme in the reliability of a system. 
Until our work, implementation tests have been the most accurate evaluation method. 
However, these tests need a long monitor test for a large amount of samples. 
For example, for the software error rate of a product, intended to evaluate the ability of several hundred samples, requires $20,000$ hours for testing \cite{renesas}. 

\section{Probabilistic Model}
\label{Modeling}

This section explains the modeling for the FUJIMI system described in Section \ref{sec:acr}.
Our objective is to build a probabilistic model for evaluating resilience strategies against noise-attack behaviors. 
The model must simulate the resilience strategy, but not cause an explosion problem \cite{Baier}.  
Furthermore, the model must be flexible for a variety of different configurations for system optimizations. 

The model is constructed with Discrete Time Markov Chains (DTMCs) as introduced 
in Section \ref{DTMC}. 
We comprehensively explore our model's multi-dimensional
parameter space by systematically varying two key
probabilistic parameters, including: 
\begin{itemize}
\item The $noise\ occurrence\ probability$:  The external environment electronic shock 
makes the $cdata$ incorrect and makes the embedded system fall into failure. 
The probability value depends on the external environment where the embedded system works. 
\item The $err\ detection\ probability$: The detection of an error in the current data is also another probability, which means that an incorrect $cdata$ will be either be detected or ignored. The probability value depends on the detection ability of the hardware. 
\end{itemize}

\begin{center}
\begin{figure}
\includegraphics[height=8cm,width=15cm]{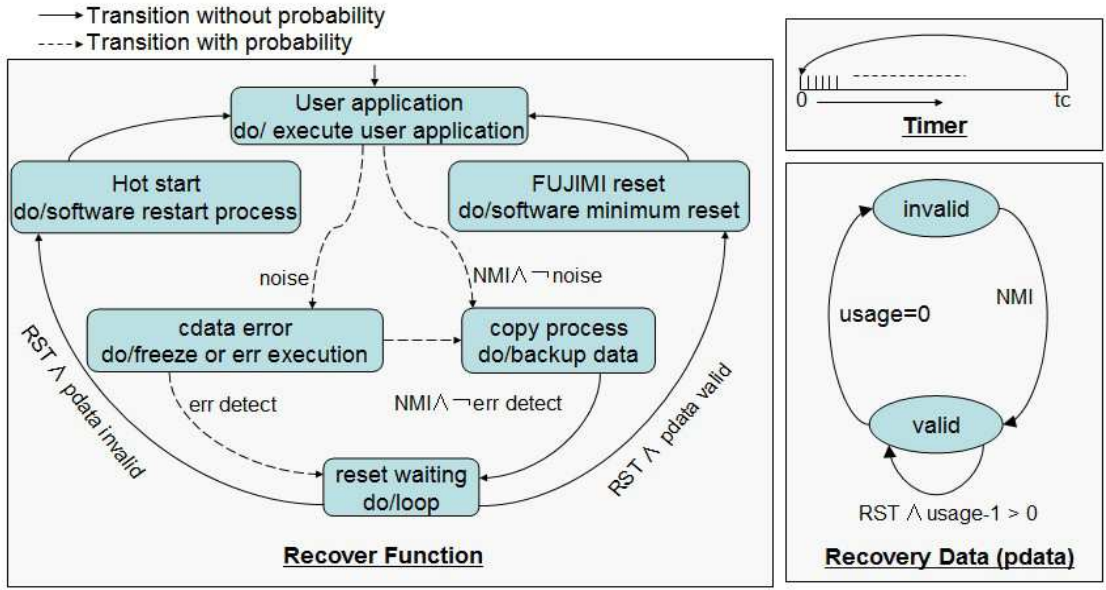}
\caption{The Model Transitions}
\label{fig:trans}
\end{figure}
\end{center}

\begin{center}
\begin{figure}
\includegraphics[height=9cm,width=16cm]{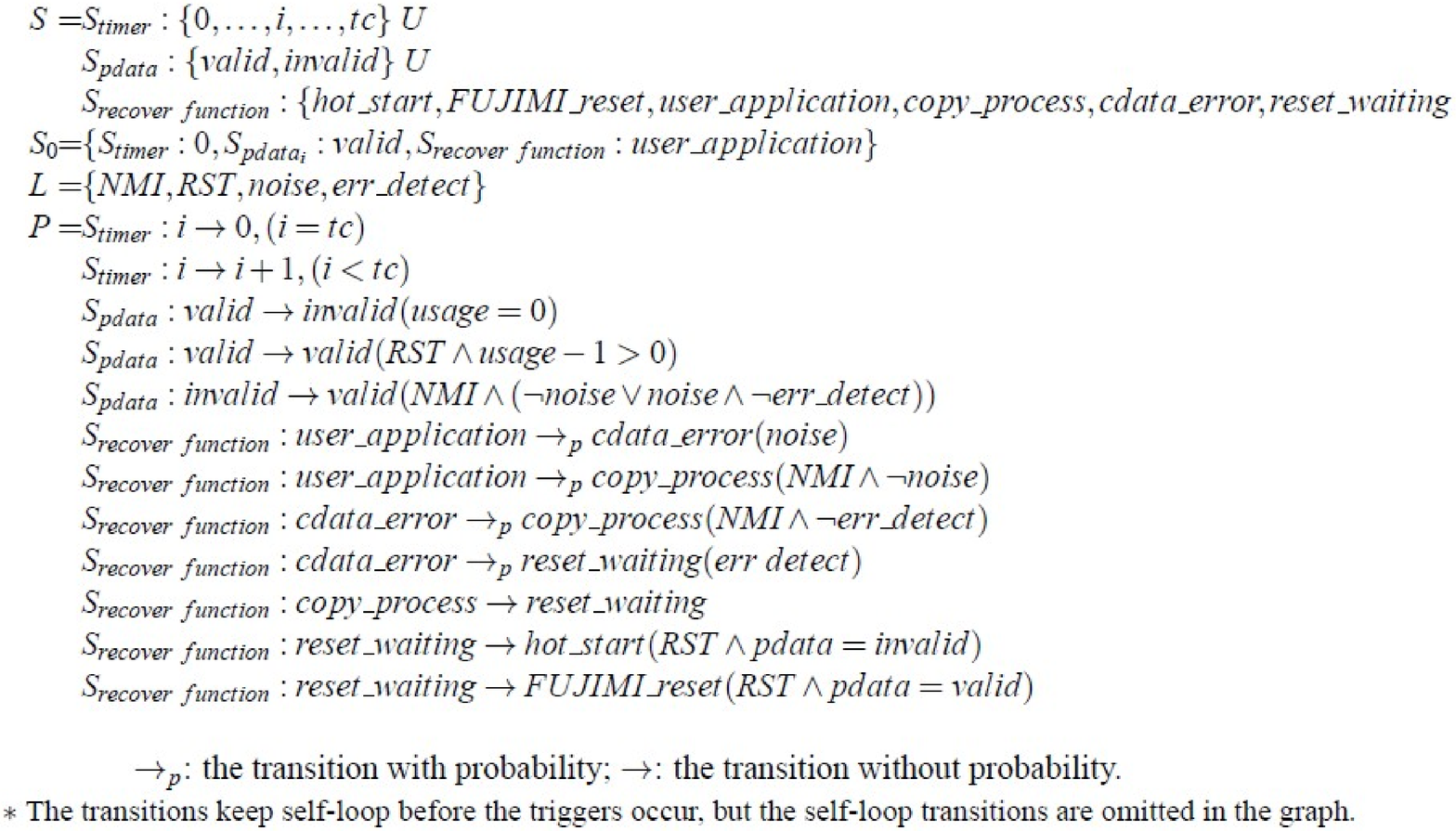}
\caption{The Model Definition}
\label{tab:modelDef}
\end{figure}
\end{center}

The model is decomposed into three modules:  
$Timer$, $Recovery\ Data$, and $Recover\ Function$ in Figure \ref{fig:trans}. 
The definition is in Figure \ref{tab:modelDef}. 
$Recovery\ Data$ and $Recover\ Function$ interact and coordinate with each other 
synchronizing with $Timer$.   
Actions are used to force two or more modules transition simultaneously \cite{PRISM}.  For example, the following codes are from Timer module and Recovery Function module respectively.  The transitions are synchronized with the action $tick$, which simulates the clock $t$ in the embedded system. 
The upper code refers that the noise occurs with the probability $prob\_noise $ when the clock ticks forward before reaching $tc$ (time cycle). 
The lower code refers that the CPU keeps normal before the clock reaches $NMI$ if the noise does not occur. \\

{\small
$[tick] (t<tc) \rightarrow  prob\_noise: (noise\prime = true) \wedge (t\prime = t+1) + (1-prob\_noise) :(noise\prime = false) \wedge (t\prime = t+1) $

$[tick] (t<NMI) \wedge (cpu = normal) \wedge (noise = false) \rightarrow (cpu\prime = normal)
$
}

\noindent 
{\bf Timer:} 
The timer has a simple loop transition as $0\rightarrow 1\rightarrow 2\dots\rightarrow tc\rightarrow 0 \rightarrow 1\rightarrow \dots$ with an assigned cycle value $tc$.  
The noise occurs with a probability for each tick according to the environments. 
The transitions are shown in Figure \ref{fig:trans} (Timer). 

\noindent 
{\bf Recovery Data ($pdata$):} 
As shown in Figure \ref{fig:trans} (Recovery Data), $pdata$ transits between $valid$ and $invalid$, according to the system status and recover strategy.
The transitions are deterministic. 
Each $pdata$ has a Boolean value that records if the data is correct, and an Integer value to record the usage times. 

At the reset timing when $RST$ signal arises, the $pdata$s transit self-loops with a decreasing usage number, and transit from $valid$ to $invalid$ when the $pdata$ exhausts to $0$.   
At the save timing when NMI is triggered, the state transits from 
$invalid$ to $valid$ and the usage is set to the initial number $m$.

\noindent 
{\bf Recover Function:} 
As shown in Figure \ref{fig:trans} (Recover Function), the recover function processes the recover algorithm for save and reset when $NMI$ and $RST$ interrupts are triggered respectively, and also processes the corresponding process when noise occurs.  
The $cdata$ and $pdata$ are operated by the recover function, while synchronizing with the $Timer$. 
The initial state is $user\_application$ representing the process when the $cold\_start$ has finished correctly. 

For the recover function, the system transits within the self-loop at the state $user\_application$ when the user application executes correctly. 
$noise$ will occur with probability, which causes $cdata$ to transit to the status $cdata\_error$. 
At the timing of $NMI$, if $cdata$ is correct or if $cdata$ is error, but is not detected, 
the system will enter into the $copy\_process$ status where the process of Figure \ref{fig:sr} (upper part) is carried out. After $copy\_process$ finishes, the system will enter into $reset\_waiting$. 
If the error status is detected, and the system will enter into $reset\_waiting$ immediately.  
In this case, the $NMI$ will be ignored and the copy process in Figure \ref{fig:sr} (upper part) will not be carried out. 

The state transits with a self-loop at the state $reset\_waiting$ 
 until the $RST$ signal arises.  
At the timing where the $RST$ exception occurs, 
the system will enter into the $FUJIMI\_reset$ where the process of Figure \ref{fig:sr} (lower part) is carried out if there is a valid $pdata$. 
But when all the $pdata$ become $invalid$ and there is no valid data for system recovery, the system transits to $hot\_start$ state. 

\section{Qualitative and Quantitative Evaluations}
\label{analysis}

PRISM supports the automated analysis of a wide range of qualitative and quantitative properties of probabilistic models with the probability temporal logic PCTL. In this section, we illustrate the qualitative and quantitative evaluations for the FUJIMI system, using the experimental data with three applications named $Sensor$, $Logger$ and $Ballon$ as shown in Table \ref{app}\footnote{$Sensor$ receives the data from the anemometer and then sends the data to a processor; $Logger$ saves the data received from the sensor to SD memory; $Balloon$ is an action game with a touch panel. The periods of time are measured through the embedded codes. The embedded codes manipulate the external terminal, and the oscilloscope records the external terminals changes.}.   
The experimental environment was as follows. 
Model checker: PRISM 4.0.3; 
CPU: Intel 1.70GHz; 2.9 GB RAM. 
The code for the model was about 130 lines. 
{\footnotesize
\begin{table}[htb]
\begin{tabular}{| l | c | c | c | c |c | c | } \hline\hline
 & $NMI\ Process$ & $RST\ Process$&$NtoR\ Duration$& $User\ Application$&$Hot\ Start$ & $Cold\  Start$ \\ \hline
Sensor    & 9.6$\mu$s  & 16.2 $\mu$s  &  500 $\mu$s & 14.9 ms    & 110 $\mu$s     & 117 $\mu$s \\ \hline 
Logger    & 94 $\mu$s  & 114 $\mu$s  & 300 $\mu$s  & 49.6 ms   & 170 ms    & 220 ms\\ \hline
Ballon     & 66 $\mu$s  & 84 $\mu$s & 25 $\mu$s     & 14.4 ms   & 76 ms     &780 ms\\ \hline
\end{tabular}
\\
\caption{Configuration of 3 applications}
\label {app}
\end{table}
}

$NMI\ Process$ is the time for the NMI process to save $pdata$. 
$RST\ Process$ is the time for FUJIMI process to reset $pdata$.
$NtoR\ Duration$ is the time from the beginning of periodical NMI to the beginning of the RST; 
$User\ Application$ represents the continuous application operation time, which must not be disturbed. 
For example, if an application communicates with the MPU, 
time for the minimum necessary time must be reserved because 
it must not be disturbed.  
$NtoR\ Duration$ is usually set to 3 times that of $NMI\ Process$ time for a sufficient process. 
The periodical cycle must be larger than the total of $NtoR\ Duration$, $RST\ Process$ and $User\ Application$. 
$Hot\ Start$ is the process time to reset when there is no $pdata$ to use. 
In this case, the system restarts from the beginning of the application. 
$Cold\ Start$ is the process time for the system to recover from failure status without the FUJIMI system. 
We provide the following evaluations for the model:   
\begin{itemize}
\item Qualitative verification: 
\begin{itemize}
\item Formal proves: We formally prove that the system satisfies the desired properties.
\end{itemize}
\item Quantitative assessment:
\begin{itemize}
\item System Configuration Optimization: We find the optimal value of the system's configurations according to different noise environments in which the embedded system works. 
\item System Failure Reduction: We give  quantitative evidence to demonstrate that system failure is reduced. 
\item Average DownTime Hours/Year: ADT is an average value, which is the time to show how long the system is out of work for 1 year. ADT is the general industry criteria. 
From our analysis, the FUJIMI system decreases the ADT (improves the effectiveness) for most of real world applications. 
\end{itemize}
\end{itemize}

\subsection{Qualitative Verification}
\label{quali}
This process formally proves that the model complies with the resilience strategy requirements. 
Qualitative properties based on Markov chains typically require certain events with probability. 
These properties can be used to verify persistence and reachability \cite{Baier}. 
We applied the following properties expressed with PCTL formulas $1-6$ (described below) to the model, 
and refined the model until the values of  formula are satisfied\footnote{The formulas are expressed in this paper for convenience, but may be different with actual PCTL formulas. For example,  $\exists i \wedge usage(pdata_i) >0$ is 
$usage_{(pdata_1)}>0 \vee usage_{(pdata_2)}>0 $ if there are two $pdata$ in the model because the operators as `$\exists$' are illegal in DTMC.}.  The model is refined until the above formulas are satisfied, that demonstrates the model comply the desired  properties. 

$(1): {\bf P}=? [{\bf F}\ {\bf G} (cdata\_error) ]$

$(2): {\bf A} [{\bf G} (usage_{(pdata_i)} < usage_{(pdata_{j})} \wedge (i < j) ] $

$(3): {\bf P}^{\ge 1}\  [{\bf F}\ hot\_start ] \hspace{55mm} $

$(4): {\bf A} [{\bf G} (reset\_waiting  
\wedge (\exists i: i>0 \wedge usage_{(pdata_i)} >0)  \wedge \neg noise)  \Rightarrow  
({\bf P}^{\ge 1} [{\bf F} (\neg cdata\_error) ]) ] $ 

$(5): {\bf A} [{\bf G} (noise \wedge reset\_waiting \wedge (\nexists i: i>0 
\wedge usage_{(pdata_i)} >0) )  \Rightarrow  ({\bf P}^{\ge 1} [{\bf F} (hot\_start) ]) ]$ 

$(6): {\bf S}=?[hot\_start]) /({\bf S}=?[cdata\_error] \leq 1 $

Formula (1) is used to inquire the probability 
 that $cdata$ can always be wrong in the future. 
This value must be $0$ for the following reasons: 
The system can be recovered with $pdata$ if there is a valid $pdata$, and can be reset initially if all the $pdata$ are invalid. 
The algorithm does not allow the wrong status of $cdata$ to continue forever.   
Formula (2) shows that the process for $RST$ is always from the most recent valid $pdata$ ($usage_{(pdata)}$ is the available usage number of $pdata$). 
This number means that on {\bf A}ll the path, the {\bf G}lobal samples satisfy that 
the $pdata$ with smaller subscript is always used before the larger subscript. 

Formula (3) means that no matter how small the probability, 
the $hot\_start$ status can be reached. 
Formula (4) can be read as: when the system enters into the loop of $reset\_waiting$ status, and there is a valid $pdata$, the system can be recovered if the noise no longer occurs. 
This property is itself resilient and can be understood easily. 
Formula (5) means that if there is no valid data for recovery and the noise occurs,  
the system will  definitely enter into $hot\_start$. 

Let ${\bf S}=?[cdata\_error]$ and ${\bf S}=?[hot\_start]$ 
be the values associated with the average of the $cdata\_error$ statistic and $hot\_start$ statistic. 
 Formula (6) implies the probability that $hot\_start$ is always smaller than the probability of $cdata\_error$ caused by $noise$. This Formula verifies that the FUJIMI system is always effective in reducing system failure. 

\subsection{Quantitative Assessment}
This section provides three kinds of quantitative assessment:  
System configuration optimization, 
system failure reduction, and 
Average DownTime.   

\subsubsection{System Optimizations}
\label{opt}

In this section three kinds of configurations are adjusted to achieve the minimum system failure or the highest effectiveness.   

\begin{figure}[htbp]
\begin{tabular}{cc}
\begin{minipage}{0.5\hsize}
\begin{center}
\hspace{-6mm}\includegraphics[height=4.5cm,width=7.5cm]{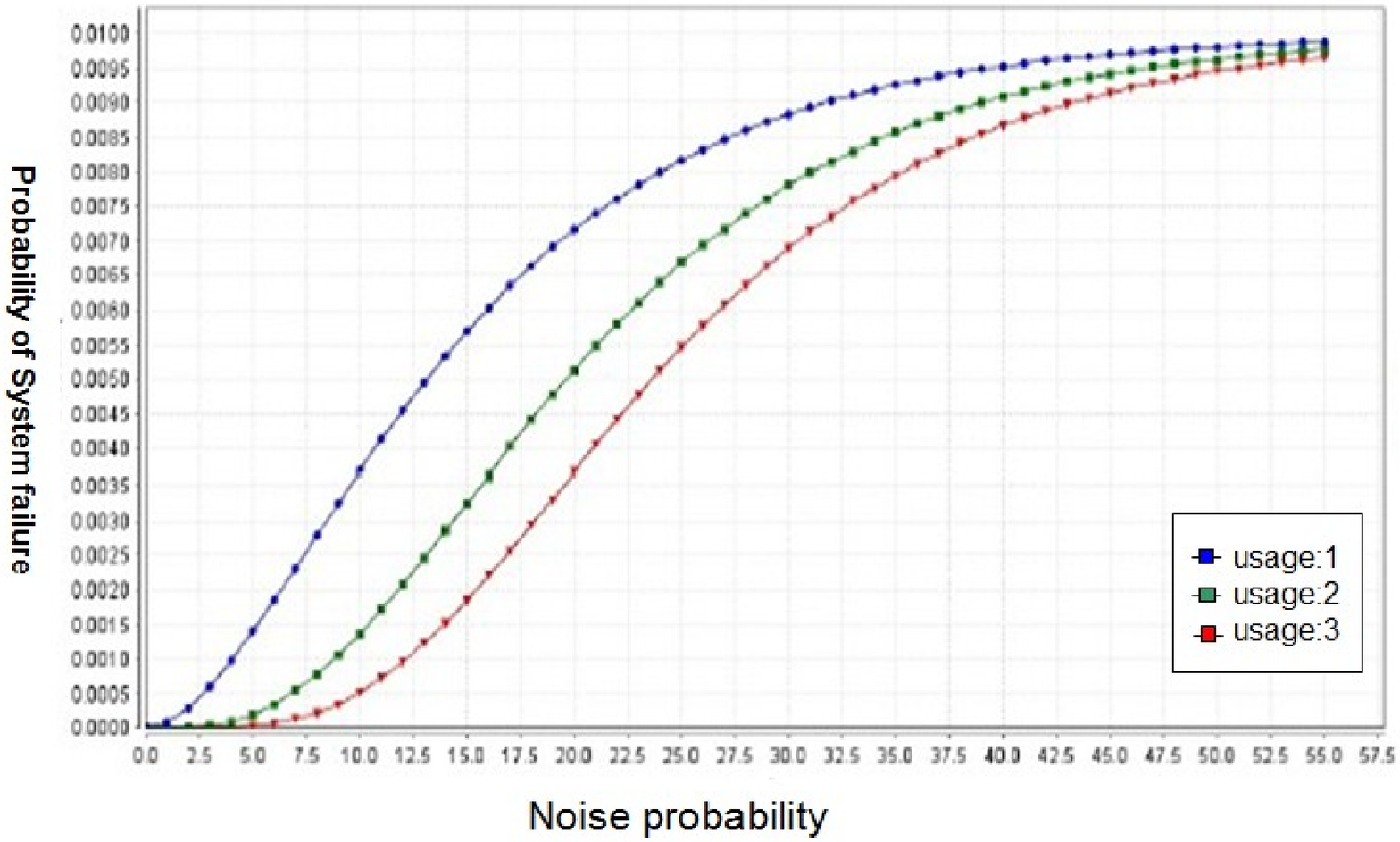}
\caption{{\footnotesize The System Failure Probability for Different Usage Times}}
\label{fig:reuse}
\end{center}
\end{minipage}
\begin{minipage}{0.5\hsize}
\begin{center}
\hspace{-6mm}\includegraphics[height=4.5cm,width=7.5cm]{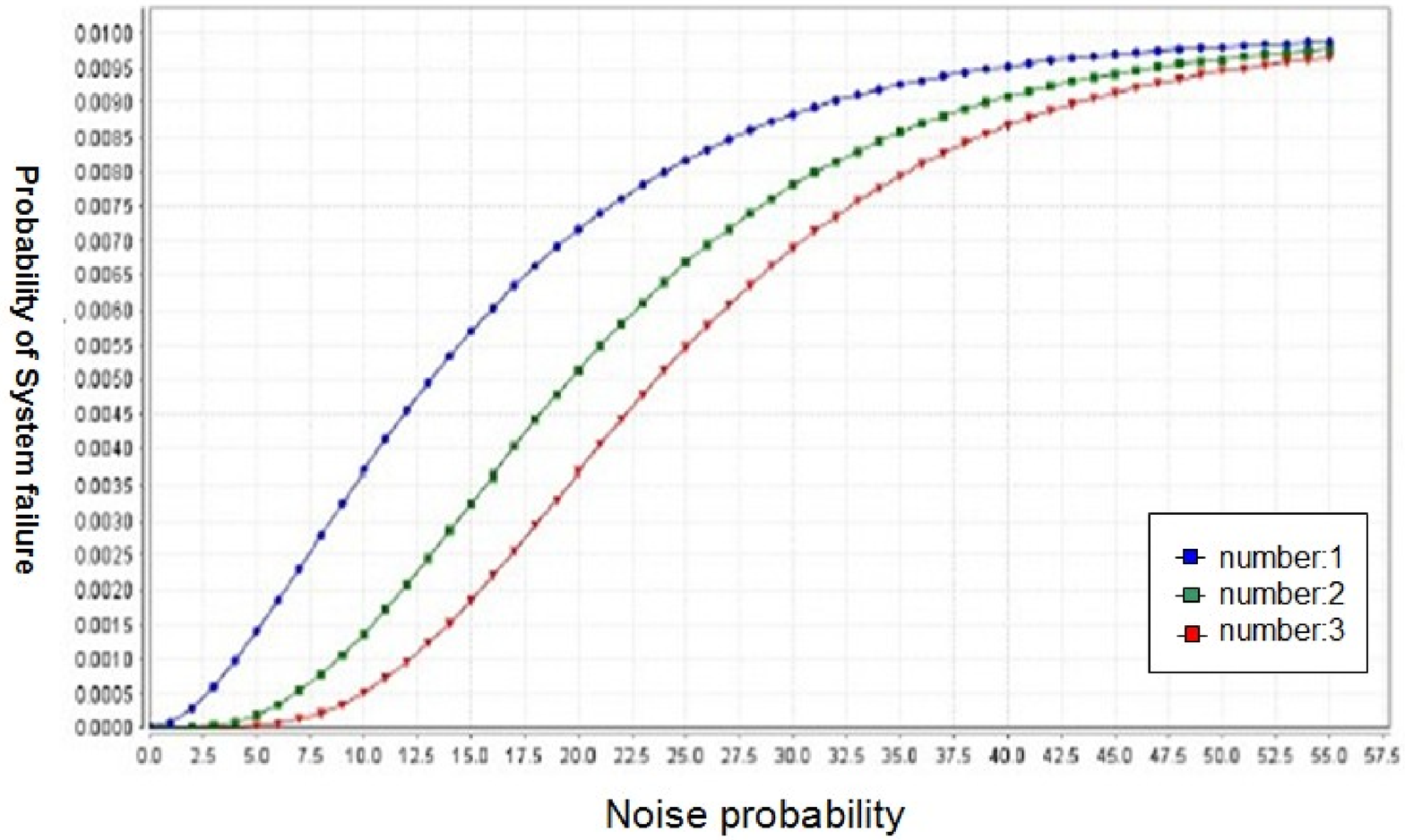}
\caption{{\footnotesize The System Failure For Different Recovery Data Numbers}}
\label{fig:buffs}
\end{center}
\end{minipage}
\end{tabular}
\end{figure}

\noindent
{\bf Usage Times for $pdata$:}
\label{sec:Opt1} 
We change the usage times of $pdata$ from 1-3 and the results are shown in Figure \ref{fig:reuse}.   
The horizontal axis is the increasing probability of noise occurrence (with a base of 10,000, for example 57.5 on the horizontal axis represents the noise probability of 57.5/10,000), and the vertical axis is the probability of system failure.
The result shows that the higher the usage time, the lower the probability of system failure.
The difference increases as the noise occurrence decreases. 
The values can help users to properly select usage times for different noise environments and the user's requirements of system failure probability.

\noindent
{\bf The Recovery Data Number:}
\label{sec:Opt2}
Another system quantity which can be optimized is the $pdata$ number. 
The number can be adjusted to achieve the most reduction in system failure. 
The probability of system failure decreases when the recovery data number increases. 
In the experiments, we sample the ${\bf S}=?[hot\_start]$ for different noise probabilities. 
The result is shown in Figure \ref{fig:buffs}.  
The horizontal axis is the increasing probability of noise occurrence (with a base of 10,000, for example 57.5 on the horizontal axis represents the noise probability of 57.5/10,000), and the vertical axis is the probability of system failure.
The system failure is reduced as the number increases, but a higher number implies memory consumption and a higher cost in time for the copying process.  

\noindent
{\bf Periodical Cycle:} 
For the resilience strategy described in Section \ref{sec:acr}, 
when the noise causes an error, the system runs into a waiting state until the reset signal $RST$ arises. 
In different environments, noise will occur frequently or sparsely. 
If the noise occurs frequently, 
then the period must be shorter, which consequently, makes the waiting time shorter. 
On the contrary, if the noise occurs sparsely, the periodical cycle must be longer to reduce the overhead of the save and reset process. 

An optimized cycle can be found with the evaluation of system effectiveness. 
For this goal, after every kind of transition is annexed with a $cost$ (called $reward$ in PRISM),  
the proportion of the execution time for user application to all the elapsed time is calculated. 
Formula (7) is the proportion of time of user application to all the elapsed time representing the effectiveness of FUJIMI system. 
The value in Formula (7) is annexed with the user application Balloon in Table \ref{app}. 
As the formula is to calculate the proportion of $weight_{available}$ and $weight_{total}$, the values in Formula (7) are 10 times smaller than the actual values in Table \ref{app} to avoid the explosion problem. 
Timer costs $1$ for $user\ application$ and $reset\ waiting$, but costs $7,600$ for $hot\ start$ for each tick.

$(7)Effectiveness={\bf R}\{``weight_{available}/weight_{total} "\}$
\begin{table}[h]
\begin{tabular}{ l | l  } 
\hspace{6mm} $rewards\ weight_{total}$  & \hspace{6mm}$rewards\ weight_{available}$ \\ 
\hspace{10mm} $user\_application: 1;$ & \hspace{10mm} $user\_application: 1;$  \\
\hspace{10mm}$reset\_waiting: 1;$ &\hspace{6mm}  $endrewards$ \\
\hspace{10mm}$hot\_start: 7,600; $ &  \\
\hspace{6mm}$endrewards$ & \\
\end{tabular}
\end{table}

\begin{figure}[htbp]
\begin{tabular}{cc}
\begin{minipage}{0.5\hsize}
\begin{center}
\includegraphics[height=4.5cm,width=7.6cm]{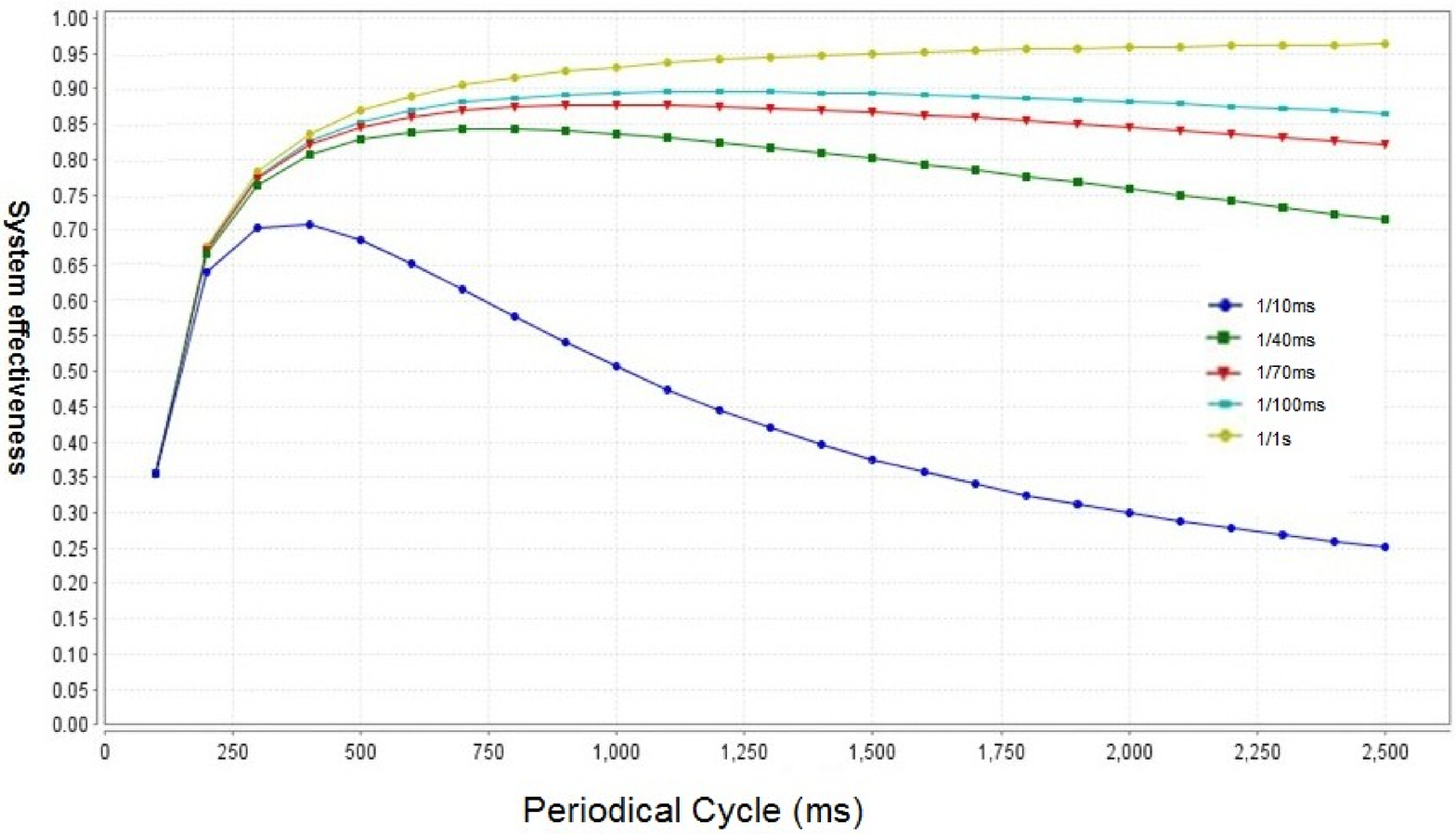}
\caption{{\small The System Effectiveness for Different Cycle Period}}
\label{fig:effect}
\end{center}
\end{minipage}
\begin{minipage}{0.5\hsize}
\begin{center}
\includegraphics[height=4.5cm,width=8cm]{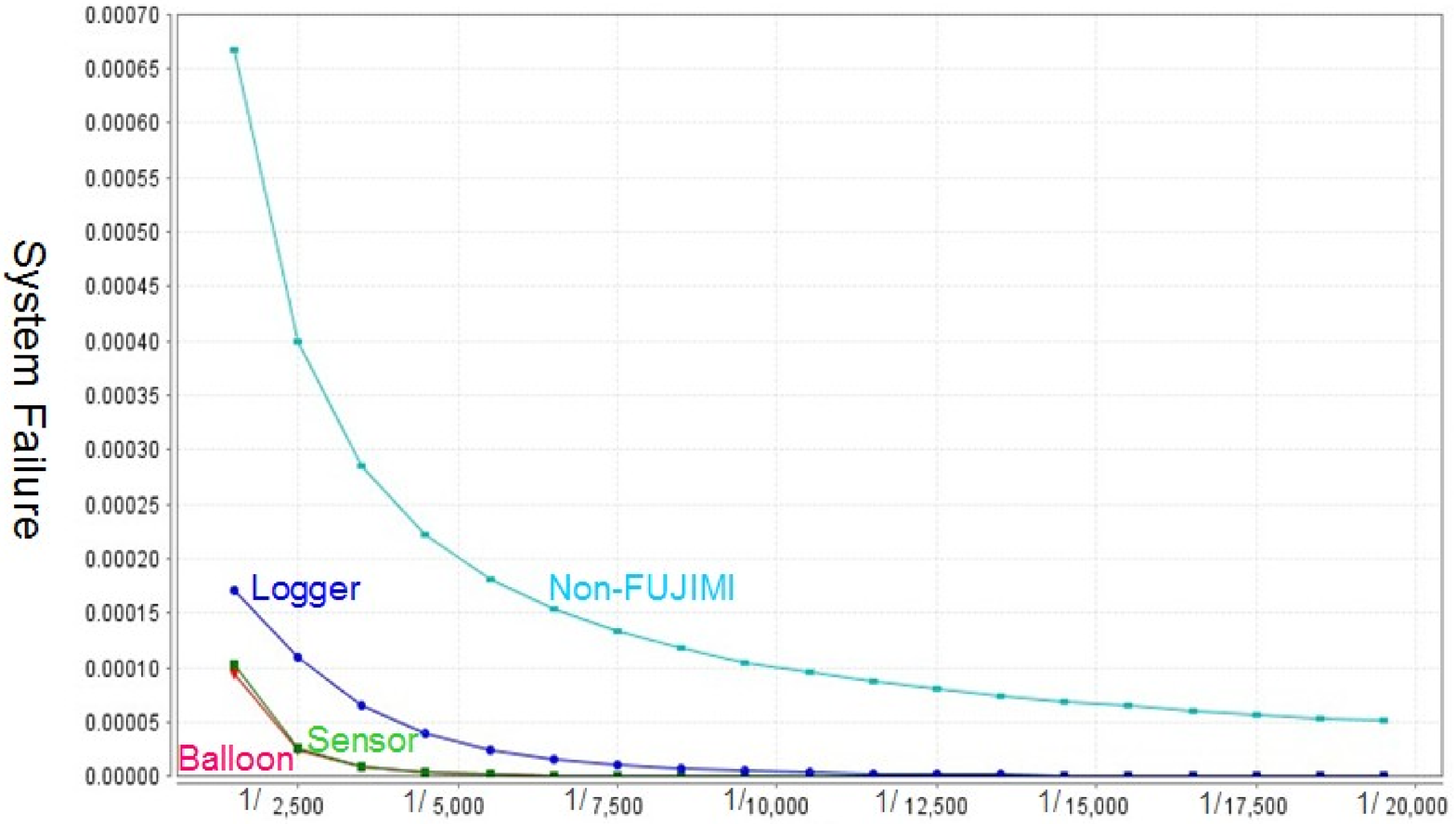}
\caption{{\small The System Failure with and without FUJIMI System}}
\label{fig:reduction}
\end{center}
\end{minipage}
\end{tabular}
\end{figure}

Figure \ref{fig:effect}  shows that the most effective time cycles are different for different probabilities of noise. The horizontal axis is the value for the cycles.  
The 5 lines represent the effectiveness for 5 kinds of noise probability of  
$1/10ms$, $1/40ms$, $1/70ms$, $1/100ms$ to $1/1s$.
The most effective point of the cycle period can be found from the graph. 
For example, the best effective point is $4ms$,$9ms$, $12ms$, $14ms$, $25ms$ for the 5 kinds of noise probability respectively. 
The results show that the periodical cycle should be set shorter if the overhead is small enough because this can avoid the $reset\ waiting$ to consume much of the effect time. However, when the noise probability is small, 
the values increase with a  slow curve after violently bending points.  
As mentioned at the beginning of this Section, the cycle must be larger than the total value of 
$User\ Application$ time, $NtoR\ Duration$ time, and $RSR\ Process$ time. 
The optimized cycle must be set to the larger one between the total value and the bending point. 

\subsubsection{System Failure Reduction}
After optimization, the system becomes deterministic, and the system failure with and without FUJIMI system can be calculated respectively. 

$(8): Failure= {\bf S} { [hot\_start]}$

The results of Figure \ref{fig:reduction} serve to demonstrate that the FUJIMI system reduces system failure caused by noise.  The horizontal axis represents the decreasing noise probability and the vertical axis represents the statistics value of the times that the system enters into $Hot\ Start$ states as shown in Formula (8).  
The most upper line represents the probability of system failure without the FUJIMI system, and the lower three lines represent the system failure of the three applications in Table \ref{app} with the FUJIMI system.  

\subsubsection{Average DownTime (ADT)} 
Average DownTime (ADT) stands for an industry a criterion that evaluates systems which must run continually.
The value is the average time in which systems malfunction and cannot work within one year. 
We use formula (9) to evaluate the ADT. 
$8,760$ is the hours in one year.
$weight_{total}$ represents the total process time, and $weight_{available}$ is the effective process time for user application. 
Cost for $hot\_start$ is substituted as $11$, $17,000$, $7,600$ for the user application $Sensor$, $Logger$, $Ballon$ respectively (10 times smaller than the actual values).
 
$(9)ADT=\ 8760*(1-{\bf R}\{``weight_{available}/weight_{totol}"\})$
\begin{table}[h]
\begin{tabular}{ l | l  } 
\hspace{6mm} $rewards\ weight_{total}$  & \hspace{6mm}$rewards\ weight_{available}$ \\ 
\hspace{10mm} $user\_application: 1;$ & \hspace{10mm} $user\_application: 1;$  \\
\hspace{10mm}$reset\_waiting: 1;$ &\hspace{6mm}  $endrewards$ \\
\hspace{10mm}$hot\_start: 7,600; $ &  \\
\hspace{6mm}$endrewards$ & \\
\end{tabular}
\end{table}

\begin{wrapfigure}{l}{8.2cm}
\includegraphics[height=4cm,width=8.2cm]{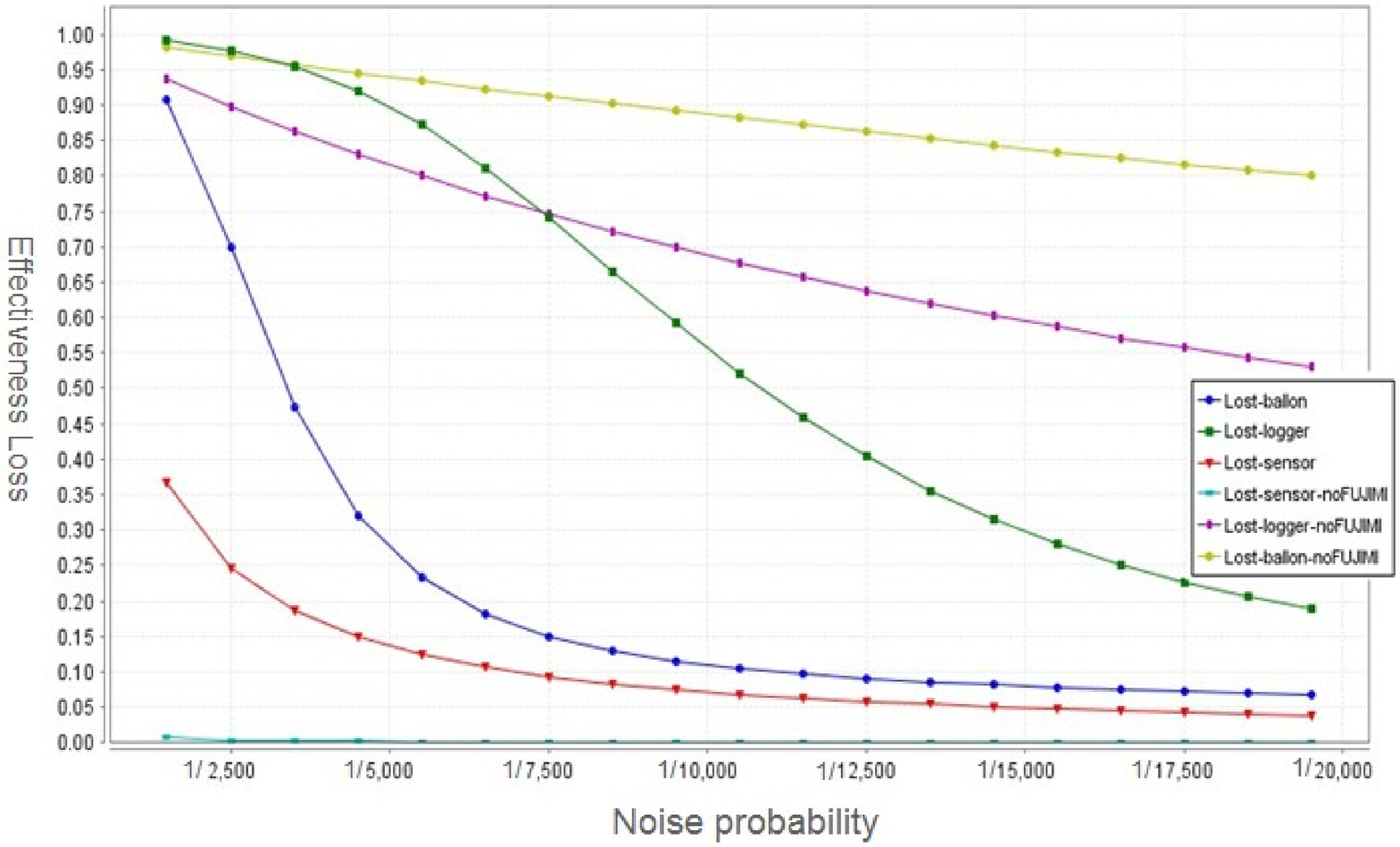}
\caption{Effectiveness Loss With and Without FUJIMI}
\label{fig:adt}
\end{wrapfigure}

Figure \ref{fig:adt} shows three pairs of lines for the applications in Table \ref{app}. Each pair indicates the values of loss proportion $(1-{\bf R}\{ ``weight_{available}/weight_{totol}" \})$ with and without the FUJIMI system when the noise probability decreases along the horizontal axis.  The restart processes without FUJIMI consume the $Cold\ Start$ time in Table \ref{app}. 

The application $Sensor$ has a light process time for the $Cold\ Start$ process.  The FUJIMI system makes the effectiveness worse because of the waiting time as well as the overhead in each cycle. For the application $Logger$, because the overhead and $Hot\ Start$ time are relatively heavy, the FUJIMI system is beneficial from the crossing point where the noise probability is $1/75,000 \mu s$. In other words, the FUJIMI system improves the system effectiveness only if the noise probability is smaller than $1/75,000 \mu s$. In the real world, the noise occurs with far smaller probability than $1/75,000 \mu s$. 
For the applications $Ballon$, the FUJIMI system brings absolute benefit as the overhead and restart time are heavy. 

The results compare the effectiveness loss with and without FUJIMI system, and indicate that the FUJIMI system makes the effectiveness worse if the process for $Cold\ Start$ is light enough. 
Nonetheless, the system without FUJIMI system is an ideal situation in that whenever the error status occurs the restart can be processed immediately, which is impossible in the real world. In general, the processes are delayed only after resilience functions as the watch-dog or manual power reset. This latency cannot be predicated. As the applications represent typical characters with the light, middle and heavy $Cold\ Start$ processes, we can affirm that the FUJIMI system can bring benefit to most real world systems.

\section{Related Work}
\label{related}
Few works have provided an analysis for temporary system failure caused by noise, which, in real world, cannot be ignored. Our target system is very original for this kind of temporary system failure, which overall improves resilience with minimum software redundancy. 
Although there are not exactly similar related works, several works for resilience systems or fault tolerance with probabilistic model checking exist.

The paper \cite{Joanne} presented a quantitative analysis to three kinds of resilience models using the Molkov reword model. This work can be considered as an elementary prototype of the kind of work described in this paper, 
but is different from our  work, in that our target resilience system is  very light N-version programming. There is no redundancy hardware, and the software alternative needs only minimum recovery data for a temporary effect from noise. 

Another paper \cite{Javier} presented an approach for the verification of self-adaptive systems that relies on probabilistic model-checking for obtaining levels of confidence regarding trustworthy service delivery when a system undergoes adaptation. An environment is stimulated to trigger system adaptation mechanisms, to collect experimental data, to trace the system undergoing adaptation, and to generate a model from the trace, and then verify the system properties. 
This work uses a quantitative assessment for a self-adaptive algorithm. 
Our work also uses a quantitative assessment, but our work provides more and different evaluations. Moreover, the model is initially constructed before evaluations in our work. 

Still another paper \cite{Dung} presented a simulation prototype for experimenting with resilience strategies for network systems.
This work supported expectations about resilience strategies and revealed strategy assumptions, unexpected emergent situations, and insights into strategy configurations.   
Evaluation of the system configurations to achieve optimized effectiveness is also a part of our work. 
But our model is different because both the evaluation object and methods are different. We formally model a light-weight redundant system with software against temporary system failure caused by noise, and to provide formal qualitative and quantitative evaluations.

\section {Summary of the Contribution and Future Work}
\label{contribution}

This paper presented qualitative and quantitative evaluations 
for a novelty resilience system named FUJIMI. 
The resilience system reduces temporary software failure caused by 
external noise with a light-weight data-backup mechanism. 
Until our work, implementation testing has been the most used method to evaluate system failure by the software error, but this method is time consuming and costly and only provides a low accuracy. 
Our method solved the difficult problem of evaluation for such kind of systems without a large testing sample.  
The formal results from the formal model were accurate and reliable. 

The qualitative evaluations verified that the FUJIMI system complies with resilience strategy, and the quantitative evaluations provided assessments of the FUJIMI system such as failure reduction and ADT. 
Moreover, the system configurations were optimized.  
Our method has been demonstrated to be very effective through three real world applications. Our systematic evaluation methods for qualitative and quantitative evaluations can be extended to other systems that have probabilistic behaviors.  

However, the evaluations only include temporary system failure on $level_4$.  
Safety Integrity Level (SIL) \cite{Smith} is widely used to define the safety properties of software. 
In the future, we plan to evaluate the FUJIMI system including the system failure  from $level_0$ to $level_5$, and provide assessment values for IEC 61508 \cite{Smith}.  

\section {Acknowledgments}

The authors would like to thank Isao Tatsuno of LETech Co., Ltd and Junzo Kenematu and Hajime Shirai of System Consultants Co., Ltd, who greatly contributed to this research.

\end{document}